\documentclass{elsart}
\usepackage[dvips]{graphicx}
\usepackage{amsmath}

\begin{document}
\title{Temperature oscillations and sound waves in hadronic matter}
\author{G. Wilk$^{a}$, Z.W\l odarczyk$^{b}$}
\address{$^{a}$National Centre for Nuclear Research,
        Department of Fundamental Research,\\ Ho\.za 69, 00-681
        Warsaw, Poland; e-mail: grzegorz.wilk@ncbj.gov.pl\\
         $^{b}$Institute of Physics, Jan Kochanowski University,\\
               \'Swi\c{e}tokrzyska 15; 25-406 Kielce, Poland; e-mail:
               zbigniew.wlodarczyk@ujk.kielce.pl\\
                  }

 {\today}

{\scriptsize Abstract: Recent high energy CERN LHC experiments on transverse momenta distributions of produced particles seem to show the existence of some (small but persistent) log-periodic oscillation in the ratios $R = \sigma_{data}\left( p_T\right)/\sigma_{fit}\left( p_T\right)$. We argue that they can provide us with so far unnoticed information on the production process, which can be interpreted as the presence of some kind of sound waves formed during the collision process in the bulk of the produced high density matter.


}

\section{Introduction}
\label{intro}

It is nowadays widely accepted that the large  transverse momentum distributions of particles observed in all large LHC experiments exhibit a quasi-power-like behavior which follows the Tsallis distribution \cite{QP1,QP2,QP3,QP4,QP5},
\begin{equation}
f\left( p_T\right) = \frac{n-1}{nT_0}\left( 1 + \frac{p_T}{nT_0}\right)^{-n},\quad n = \frac{1}{q-1}, \label{TD}
\end{equation}
where $q$ is the so-called nonextensivity parameter and $T_0$ is a scale factor \footnote{In what follows we use units with $c=k_B=h=1$.}. Such power-like distributions are ubiquitous and follow from different, apparently not connected, dynamical mechanisms examples of which are, for completeness,  listed and shortly discussed in Appendix \ref{sec:A}. However, not so obvious is the origin of the apparent observation that, under closer inspection this behavior is additionally adorned with some log-periodic oscillations \cite{WWln}. Although this is a rather subtle effect (seen essentially only in the ratios of the measured cross-sections to their phenomenological power-like fits, $R = f_{data}\left( p_T\right)/f _{fit}\left( p_T\right)$ ), it shows itself in all known results (i.e., in all experiments and at all energies) provided that the range of transverse momenta observed is large enough (reaching $200$ GeV). Furthermore, these oscillations cannot be erased by any reasonable change of fitting parameters. It is therefore rather unlikely that they can be attributed to some artifacts of the measurement process or to the specific details of the detector used, the more so, that they become stronger in reactions with nuclei where they grow with the increasing centrality of the collision \cite{RWWlnA}. In fact, such log-periodic oscillations are rather ubiquitous in other branches of physics whenever one deals with power-like distributions. They are usually attributed to a discrete scale invariance (connected with a possible fractal structure of the process under consideration) and are described by introducing a complex power index \cite{Sornette}\footnote{Actually, in the domain of multiparticle production processes there was another proposition \cite{SoftHard1} (cf., also \cite{SoftHard2}) based on two component, {\it soft+hard}, model of particle production. However, as was demonstrated in \cite{RWWlnA} this approach does not show any continuation of the log-periodic oscillations for large transverse momenta, the beginning of which seems to be seen in data.}.

However, the Tsallis distribution (\ref{TD}) is not a simple, one parameter power-like distribution. It has a quasi-power-like form with two parameters: the power index $n$ (connected with the nonextensivity parameter $q$) and the scale factor $T$ (becoming in the $q\rightarrow 1$ limit the usual temperature). Therefore each of them (or both) can be used to describe the observed log-periodic oscillations. The case when $n$ becomes complex for a Tsallis distribution has been discussed by us in \cite{WWln,RWWlnA,WW_E,WW_APPB,WW_ChSF}. Such an approach leads to a number of observed consequences, like a complex heat capacity of the system, complex probability and complex multiplicative noise, discussed briefly in \cite{WW_E}. The other possibility, discussed by us in \cite{WW_APPB,WW_ChSF} is to keep the power index $n$ real, but to allow for some specific log-periodic oscillations of the scale parameter $T$. Note that whereas each of the proposed approaches are based on a different dynamical picture, nevertheless, as demonstrated in Appendix \ref{sec:B}, they are numerically equivalent.

\begin{figure}[t]
\begin{center}
\resizebox{0.75\textwidth}{!}{%
  \includegraphics{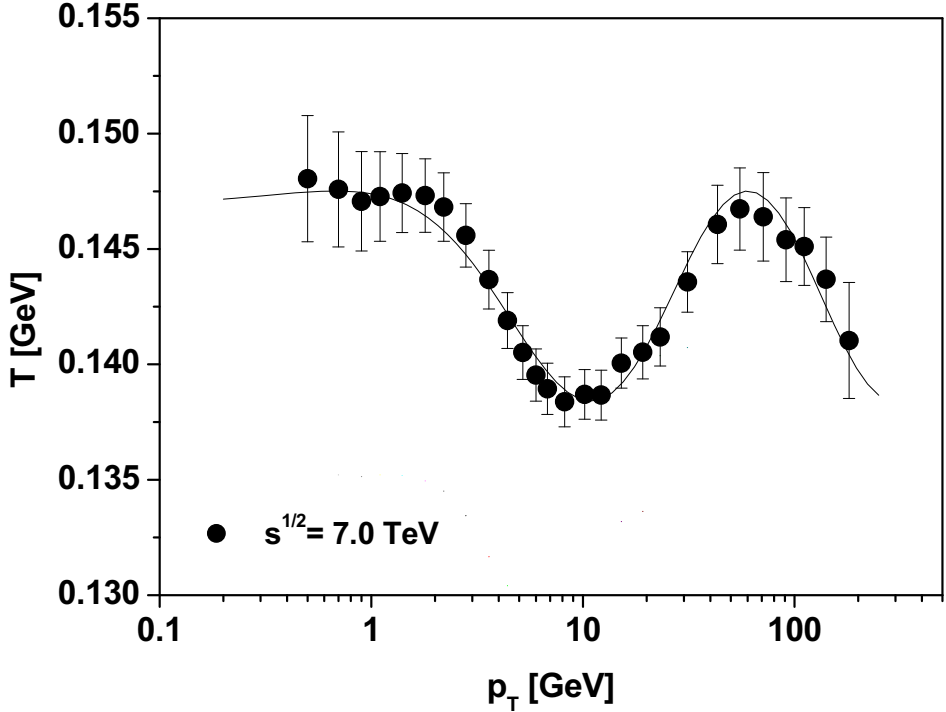}
}
\caption{ Example of the observed log-periodicity of the scale parameter $T$ (deduced from CMS data \cite{QP1,QP2}, see \cite{WW_APPB,WW_ChSF} for details).}
\label{FF1}
\end{center}
\end{figure}

In this work we continue our previous investigations \cite{WW_ChSF} in which the observed transverse momenta distributions presented in Fig. \ref{FF1} were described  using the oscillating (as function of transverse momentum $p_T$) temperature $T$ (keeping the power index $n$ constant):
\begin{equation}
T\left( p_T\right) = a + b \sin\left[ c \ln\left( p_T + d\right) + e \right]. \label{T_p_T}
\end{equation}
The origin of such oscillations (and, respectively, the meaning of the parameters $a$, $b$, $c$, $d$ and $e$ in Eq. (\ref{T_p_T})) comes from the possible energy dependence of the noise defining the corresponding stochastic processes in the well known stochastic equation for the evolution of temperature $T$:
\begin{equation}
\frac{dT}{dt} + \frac{1}{\tau} T + \xi\left( t,p_T\right) = \Phi. \label{LE}
\end{equation}
Here $\tau$ is the relaxation time, $\xi\left(t,p_T\right)$ is the $p_T$-dependent noise and it is additionally assumed that we have time-dependent observed transverse momentum, $p_T = p_T(t)$, satisfying in the scenario of the preferential growth of networks \cite{PrefGrowth} equation
\begin{equation}
\frac{dp_T}{dt} = \frac{1}{\tau_0}\left( \frac{p_T}{n} \pm T\right) \label{PrefG}
\end{equation}
(where $n$ coincides with the power index in Eq. (\ref{TD}); $\tau_0$ is some characteristic time step). The stationary dependence of $T\left( p_T\right)$, Eq. (\ref{T_p_T}), is obtained in two cases (cf., Section 3.2 in \cite{WW_ChSF}): either the noise term increases logarithmically with transverse momentum and one keeps the relaxation time $\tau$ constant:
\begin{equation}
\xi\left(t, p_T\right) = \xi_0(t) + \frac{\omega^2}{n}\ln \left( p_T\right), \label{p_T_increase}
\end{equation}
or, equivalently, one keeps the $p_T$-independent white noise, $\xi\left( t, p_T\right) = \xi_0(t)$, but allows for the relaxation time becoming $p_T$-dependent, for example assuming that
\begin{equation}
\tau = \tau\left( p_T\right) = \frac{n\tau_0}{n + \omega^2 \ln\left( p_T\right)} \label{Tau_dep}
\end{equation}
(in both cases $\omega$ is some new parameter, cf. \cite{WW_APPB,WW_ChSF} for details). It turns out that to fit data it is enough to allow for a rather small admixture of the stochastic processes with noise depending on the transverse momentum (defined by the ratio $b/a \sim 3\%$). The main contribution comes from the usual energy-independent Gaussian white noise.

Some remarks are in order here. Unlike the previous case of complex $n$, the situation of varying scale parameter (temperature) $T$ is not so widely known. The encountered here oscillatory  character of $T$ seems to be not so exotic either. Namely, note that the popular approach to strongly interacting systems, both at low and high energies and also in dense stars, is provided by hydrodynamical models. They became increasingly sophisticated and received continuous support from experimental data (especially from the observation of elliptic flow, phenomenon not so easy to explain in other approaches, which are more successful in description of distributions of the measured transverse momenta). In fact, there is compelling evidence that at heavy ion collisions at RHIC and LHC experiments we observe a kind of “perfect fluid”.  The above reasoning would be reinforced by the observation of the production of waves in a hadronic medium. They could arise because in the initial phase of the collision the highly energetic partons are created, which subsequently lose their energy either by exciting modes of the medium ({\it collisional energy loss}) or by radiating gluons ({\it radiative energy loss}). In the case of a weakly coupled medium, thermalization of deposited energy occurs through a cascade of collisions among quarks and gluons in the quark-gluon plasma; in a strongly coupled medium the energy is dissipated directly into thermal excitations and sound waves. In fact, there are already some indications that such waves have indeed been formed . In relativistic heavy ion collisions we may also have hard parton-parton collisions in which the outcoming partons have to traverse the surrounding fluid before escaping and forming jets. Therefore they may form Mach shock waves during such passages, which will affect the transverse momentum distribution of the observed final particles \cite{HydroWaves}. In this work we therefore study the propagation of such sound waves in hadronic matter. More specifically, we consider the propagation of perturbations in the temperature and demonstrate that they generate log-periodic waves. Using now their Fourier transforms we intend to gain a new information about the collision process.

\section{Fourier transform picture of oscillating $T$}
\label{Fourier}

So far we have concentrated on the transverse momentum dependence of the temperature $T$ as deduced from the data, $T=T\left( p_T\right)$. But, as recently demonstrated in \cite{Hindusi}, using the Fourier transformation of such dependencies one can gain insight into the space-time structure of the collision process. It is therefore tempting to apply a similar approach in our case. It must be stressed from the beginning that whereas in \cite{Hindusi} investigations were concentrated on the possible fluctuations of $T$, we have at our disposal their additional log-periodic oscillations. The intrinsic oscillations of $T$ investigated in \cite{Hindusi} are, in our case, accounted for by the use, as our starting point, of a Tsallis distribution.  As was shown in many cases (cf., for example, \cite{WWq}) this distribution emerges from the usual Boltzmann distribution when one fluctuates the scale parameter (temperature) according to a gamma distribution. In \cite{Hindusi} the non-equilibrated systems were instead described using the Boltzmann transport equation and their space variations were investigated by means of the Fourier transformations. In this approach the log-periodic oscillations are out of reach.

\begin{figure*}[t]
\resizebox{0.5\textwidth}{!}{%
  \includegraphics{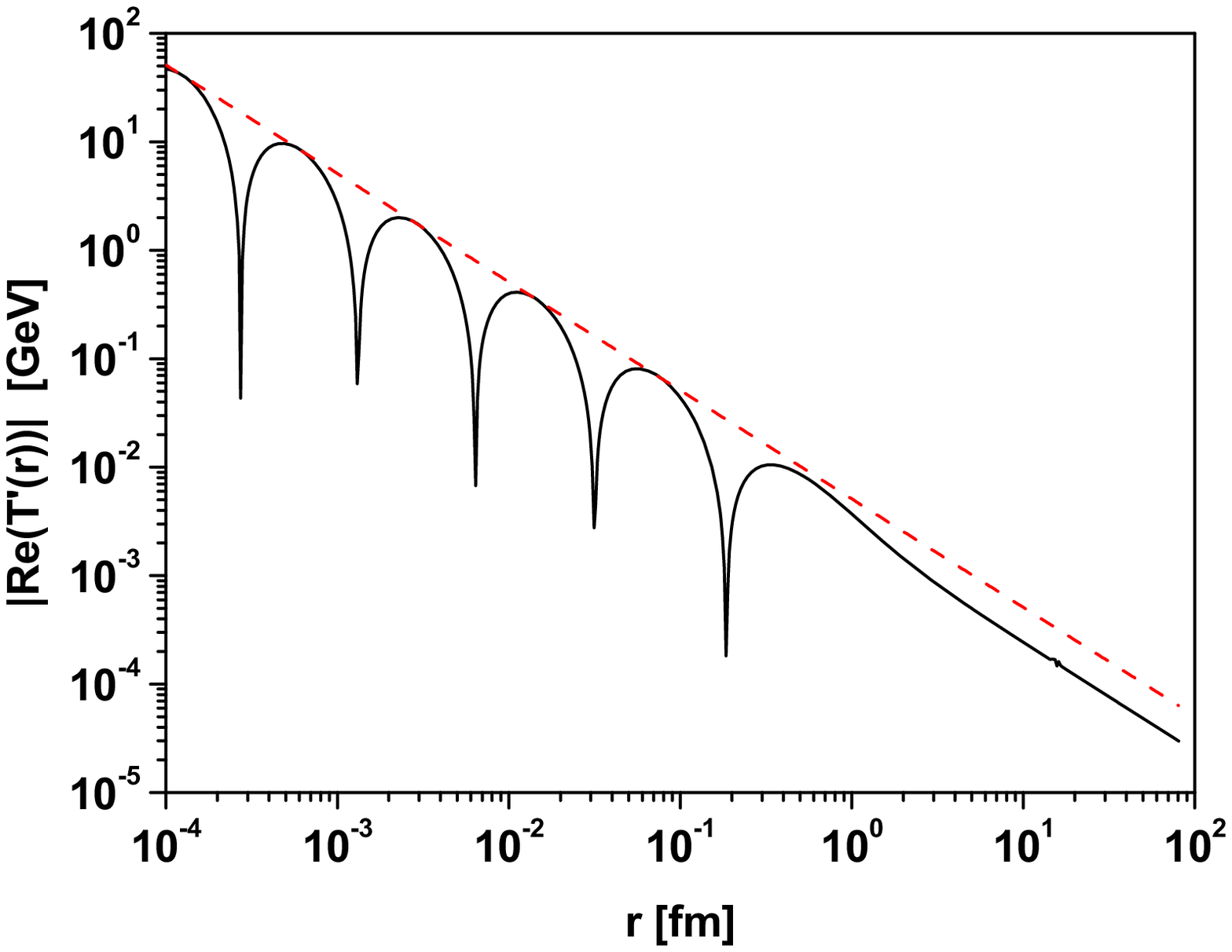}
}
\resizebox{0.5\textwidth}{!}{%
  \includegraphics{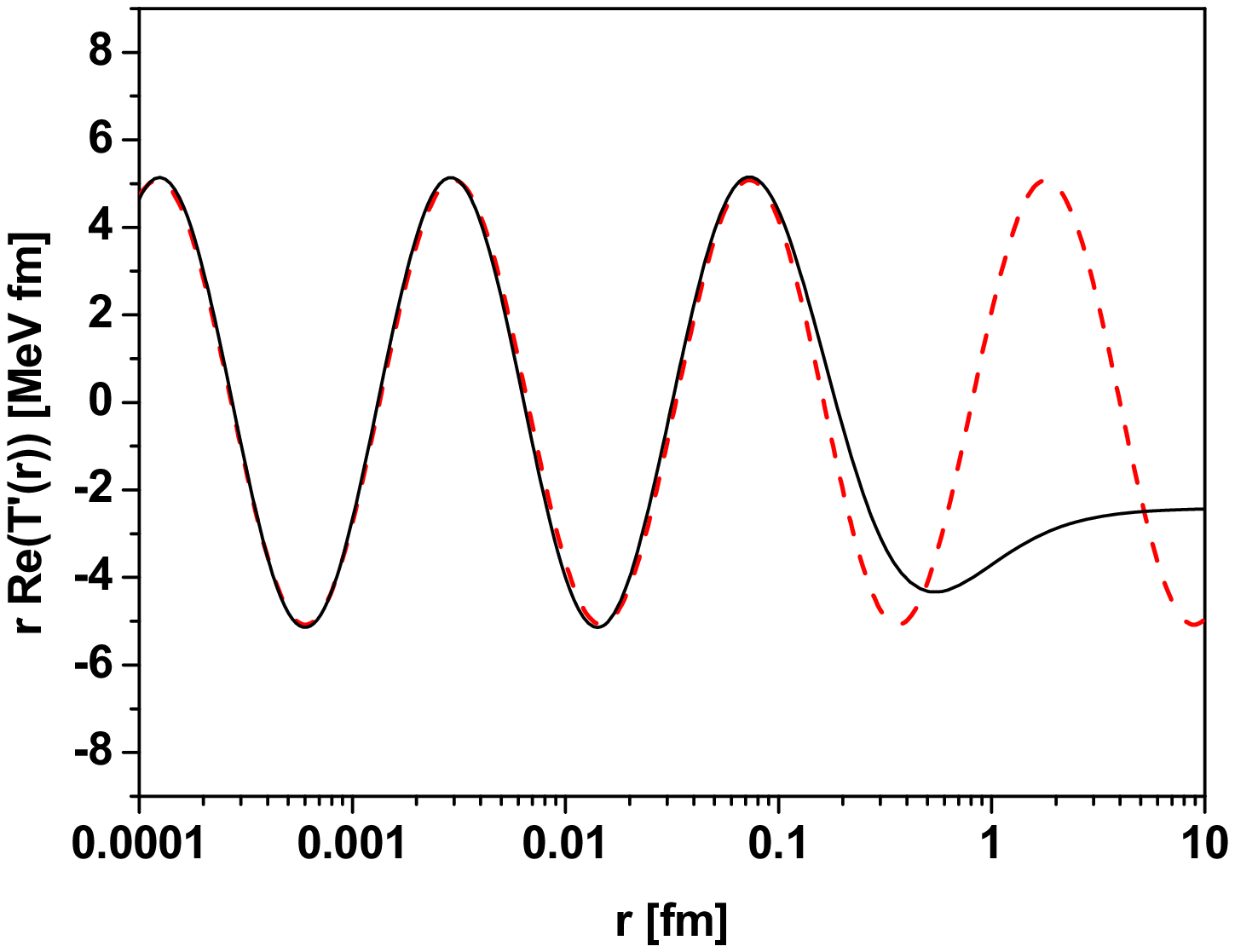}
}
\caption{ (Color online) The results of the Fourier transform of $T\left(p_T\right)$ from Eq. (\ref{T_p_T}) describing the results inferred from the CMS data \cite{QP1,QP2}. In the left panel we have $T'(r)$ (continuous line) as a function of $r$ confronted with $T'(r) = 0.0051/r$ dependence (dashed line).  In the right panel the continuous line represents $r T'(r)$ versus $r$ and the dashed line denotes the function $rT'(r) = 5.1 \sin [(2\pi)/3.2 \ln(1.24 r)]$ fitting it for small values of $r$.  }
\label{T12}
\end{figure*}

In what follows we investigate therefore the Fourier transform of log-normal oscillations of $T\left( p_T\right)$ apparently seen in data and described by Eq. (\ref{T_p_T}):
\begin{equation}
T(r) = \sqrt{\frac{2}{\pi}}\int^{\infty}_0 \, T\left( p_T\right) e^{ip_T r} dp_T   \label{FT}
\end{equation}
(note that $r$ is defined in the plane perpendicular to the collision axis and located at the collision point and denotes the distance from the collision axis). It is shown in Fig. \ref{T12}, and, as one can see, it represents some log-periodic acoustic wave forming in the source (actually, in the above mentioned plane located at the collision point). To illustrate how it can appear let us consider, as an example, the case of a cylindrical source and assume that $T(r)$ consists of a constant term, $T_0$, and some oscillating addition, $T'(r)$:
\begin{equation}
 T(r) = T_0 + T'(r)\qquad {\rm where}\qquad T' = \left( \frac{\partial T}{\partial P}\right)_S P'.  \label{oscillatingT}
\end{equation}
For small oscillations the pressure $P$ and density $\rho$  can be written, respectively, as
\begin{equation}
 P = P_0 + P',\quad\quad \rho = \rho_0 + \rho', \label{Prho}
\end{equation}
where  $P_0$  and  $\rho_0$  are the pressure and density in equilibrium. It can be shown that $T'(r)$ is related to the velocity $v$ of a sound wave \cite{Landau}. Using the thermodynamic relation
\begin{equation}
\left( \frac{\partial T}{\partial P }\right)_S = \frac{T}{c_P}\left( \frac{\partial V}{\partial T} \right)_P \label{thR}
\end{equation}
we have that
\begin{equation}
T' = \frac{c \kappa T}{c_P} v, \label{thR}
\end{equation}
where
\begin{equation}
\kappa = \frac{1}{V}\left( \frac{\partial V}{\partial T}\right)_P\quad {\rm and}\quad c = \sqrt{\left( \frac{\partial P}{\partial \rho}\right)_S} \label{kapparho}
\end{equation}
are, respectively, the coefficient of thermal expansion and the velocity of sound (note that $v << c$). Introducing the velocity potential,
\begin{equation}
\overrightarrow{v} = \overrightarrow{\nabla}(f), \label{velgrad}
\end{equation}
for a cylindrical wave we obtain the following wave equation:
\begin{equation}
\frac{1}{r} \frac{\partial}{\partial r}\left( r \frac{\partial f}{\partial r}\right) - \frac{1}{c^2}\frac{\partial^2 f}{\partial t^2} =0. \label{partialwe}
\end{equation}
In the case of a monochromatic wave, when $f(r,t) =
 f(r)\exp(-i\omega t)$, we have that
\begin{equation}
\frac{\partial^2 f(r)}{\partial r^2} + \frac{1}{r} \frac{\partial f(r)}{\partial r} + K^2 f(r) = 0, \label{eqn}
\end{equation}
where
\begin{equation}
K=K(r) = \frac{\omega}{c(r)} \label{K}
\end{equation}
is the wave number which in inhomogeneous media depends on $r$. For the wave number given by
\begin{equation}
K(r) = \frac{\alpha}{r} \label{example}
\end{equation}
the solution of Eq. (\ref{eqn}) is given by a log-periodic oscillation in the form
\begin{equation}
f(r) \propto \sin [ \alpha \ln(r)]. \label{solution}
\end{equation}
Because from Eq. (\ref{velgrad}) we have $f(r) \propto v r$, therefore, using Eq. (\ref{thR}), we can write that
\begin{equation}
rT'(r) \propto \frac{c\kappa T_0}{c_P}f(r) = \frac{c \kappa T_0}{c_P} \sin[ \alpha \ln (r)]. \label{final}
\end{equation}
This is the solution we have used in describing the $T'(r)$ deduced from data and presented in Fig. \ref{T12}.

Actually, one can look at this problem from another point of view. Namely, according to \cite{BZ1,BZ2,B1,B2},  Eq. (\ref{eqn}) with the wave number (\ref{example}) has a so-called self similar solution of the second kind. It is known from other branches of physics and is connected with the so called intermediate asymptotic encountered in phenomena which do not depend on the initial conditions because sufficient time has already passed, although the system considered has not yet reached the equilibrium state \cite{BZ1,BZ2,B1,B2}. Therefore, introducing the variable
\begin{equation}
\xi = \ln r \label{ksi}
\end{equation}
we find that Eq. (\ref{eqn}) for the wave number (\ref{example}) leads to the traveling wave equation,
\begin{equation}
\frac{\partial^2 F(\xi)}{\partial \xi^2} + \alpha^2 F(\xi) = 0, \label{TWE}
\end{equation}
which has solution
\begin{equation}
F(\xi) \propto \sin(\alpha \xi). \label{TWS}
\end{equation}

The space picture of the collision (in the plain perpendicular to the collision axis and located at the collision point) presented in Fig. \ref{T12} shows us the existence of some regular (on the logarithmic scale) structure for small distances. It starts to weaken quite early (at $ r \sim 0.1$ fm) and essentially disappears when $r$ reaches the dimension of the nucleon, i.e., for $r \sim 1$ fm.

\vspace{3mm}
\begin{table}[h]
\caption{Values of fit parameters in Eq. (\ref{T_p_T}). }\label{Table1}
\vspace{1mm}
\begin{tabular}{|c|r|r|r|r|r|}
\hline
$Collision$ & $a$ & $b$ & $c$ & $d$ & $e$ \\
\hline
$p+p$ & 0.143 & 0.0045 & 2.0 & 2.0 & -0.4 \\
$Pb+Pb$ & 0.131 & 0.017 & 1.7 & 0.05 & 0.98 \\
\hline
\end{tabular}
\end{table}
\vspace{3mm}

\begin{figure}[t]
\begin{center}
\resizebox{0.75\textwidth}{!}{%
  \includegraphics{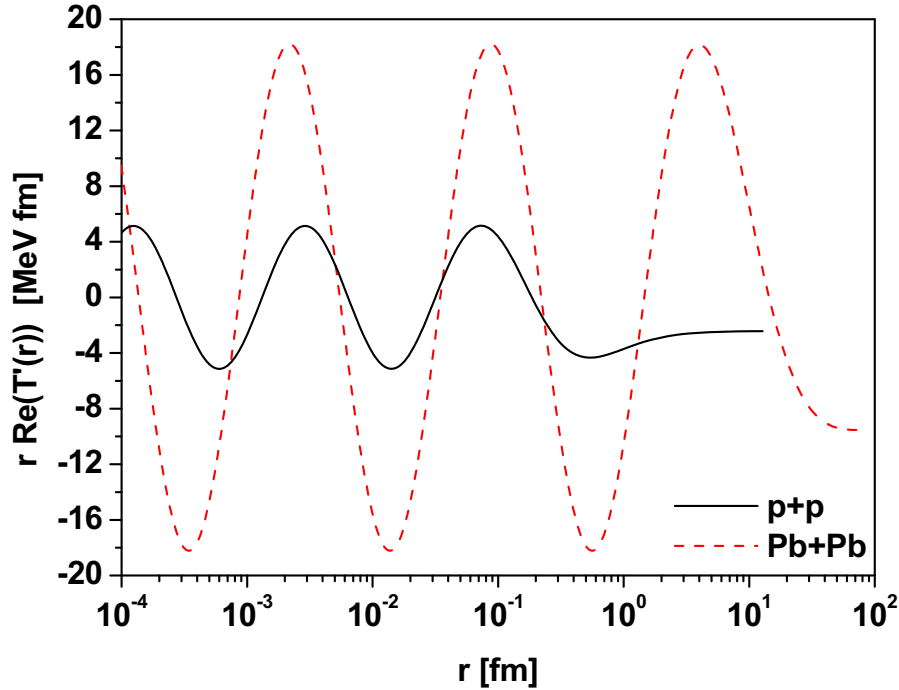}
}
\caption{(Color online) Oscillating behaviour of $r T'(r)$ for $p+p$ collisions at $7$ TeV (right-hand panel of Fig. \ref{T12}) in comparison with similar results for the most central $Pb+Pb$ collisions at $2.76$ TeV (inferred from the ALICE data \cite{QP4,QP5}).
}
\label{Pb_p}
\end{center}
\end{figure}

To shed more light on such a picture we compare in Fig. \ref{Pb_p} the oscillating behavior of $rT'(r)$ for $p+p$ collisions (as presented above in the right-hand panel of Fig. \ref{T12}) with similar results obtained for the most central $Pb+Pb$ collisions deduced from the ALICE data \cite{QP4,QP5} at $2.76$ TeV \cite{RWWlnA}.  The parameters used in both fits are listed in Table \ref{Table1}. For central $Pb+Pb$ collisions we observe $\sim 3.6$ bigger amplitude and $\sim 1.15$ longer period of oscillations.  With decreasing centrality the amplitude decreases smoothly reaching practically the same value as for $p+p$ collisions \cite{RWWlnA}. With the parameters used we have, in the region of regular oscillations, that
\begin{eqnarray}
\!\!\!\!\!(p+p):&& rT'(r) = 5.1 \sin \left[ \frac{2\pi}{3.2}\ln(r) + 0.42\right]; \label{R_Pb}\\
\!\!\!\!\!(Pb+Pb):&& rT'(r) = 18.53 \sin \left[ \frac{2\pi}{3.68}\ln(r) - 0.51\right]. \label{R_p}
\end{eqnarray}

Note that a longer period of oscillations in $Pb+Pb$ collisions now means a smaller value of the parameter $\alpha$ in Eq. (\ref{solution}) in $p+p$ collisions. Further this means that, when looking at Eqs. (\ref{K}) and (\ref{example}), because $\omega/c(r)=\alpha/r$, the velocity of sound, $c(r) = (\omega/\alpha)r$, is greater in the nuclear environment  (for $Pb+Pb$) than in $p+p$. This, in turn, means that the refractive index $n(r) = c_0/c(r)$ at position $r$ in nuclear collisions is smaller than in collisions of protons. Nevertheless, in both cases we encounter an inhomogeneous medium with $r$-dependent  $c(r)$ and $n(r)$.

These findings can be confronted with the following experimental observations. In nuclear collisions one really observes a higher speed of sound as demonstrated by the NA61/SHINE collaboration at SPS energies \cite{NA61,GGS}\footnote{~~~~~What is observed there is the velocity of sound being a parameter in the equation of state of hadronic matter described by a hydrodynamical model.}. That this should, in fact, be expected, can be visualized by considering the connection of the isothermal compressibility  of the nuclear matter,
\begin{equation}
\kappa_T = - \frac{1}{V}\left( \frac{\partial V}{\partial P}\right)_T, \label{IC}
\end{equation}
and fluctuations of the multiplicity of produced secondaries represented by the relative variance, $\varpi$,
of multiplicity fluctuations, namely \cite{Hill,Balescu}
\begin{equation}
T\kappa_T \rho_0 = \frac{\langle N^2 \rangle - \langle N\rangle^2}{\langle N\rangle} = \varpi \label{IC-NF}
\end{equation}
(where $\rho_0 = \langle N\rangle/V$ denotes the equilibrium density for $N$ particles with mass $m$ located in volume $V$). This allows the velocity of sound to be expressed by fluctuations of multiplicity because one has that
\begin{equation}
c = \sqrt{\frac{\gamma}{\kappa_T \rho_0 m}} = \sqrt{\frac{\gamma T}{\varpi m}}\quad{\rm where}\quad \gamma = \frac{c_P}{c_V}. \label{SV}
\end{equation}
Note that higher velocity of sound corresponds to lower fluctuations of multiplicity. From the data presented above one obtains that
\begin{equation}
\frac{c_{Pb+Pb}}{c_{p+p}} \simeq 1,15\quad{\rm and}\quad \frac{\varpi_{p+p}}{\varpi_{Pb+Pb}} \simeq 1.32 \label{results}
\end{equation}
which agrees with the recently obtained value for negatively charged hadrons, $\varpi_{p+p}/\varpi_{Pb+Pb} \simeq 1.29\pm0.04$ \cite{RecentNA61}.

These results can be connected with the pair correlation function, $g^{(2)}$, because the scaled variance can be written as \cite{FS}
\begin{equation}
\rho_0 \kappa_T T = 1 + \rho_0 \int d \vec{r} \left[ g^{(2)}(r) - 1\right]. \label{Corrrel_g}
\end{equation}
As shown in \cite{RW}, for central nuclear collisions the number of binary collisions exceeds that of wounded nucleons (each nucleon participates in a number of collisions with other nucleons). This results in the correlation function becoming negative which, in turn, leads to a diminishing of fluctuations of multiplicity (because the variance of the total multiplicity from a number of particular collisions is smaller that the sum of the variances of independent nucleon-nucleon collisions).

\section{Summary}
\label{Sum}

Inspired by the observation that Fourier transforms of any fluctuations (or oscillations) of the temperature $T\left(p_T\right)$ deduced from the experimental data on multiparticle production could be regarded as a source of a space-time structure of the collision process, we have used it to investigate the log-periodic oscillations in $T$ apparently observed in such data \cite{WWln,RWWlnA,WW_E,WW_APPB,WW_ChSF}. We have found that the Fourier transform picture of log-periodically oscillating $T$ represents some log-periodic acoustic wave forming in the source. Further inspection showed that the corresponding wave equation has self-similar solutions of the second kind connected with the so called intermediate asymptotic observed in phenomena which do not depend on the initial conditions because sufficient time has already passed, although the system considered is still out of equilibrium  \cite{BZ1,BZ2,B1,B2}.

When applied to $p+p$ and $Pb+Pb$ collisions our results show that in both cases one deals with an inhomogeneous medium with both density and velocity of sound depending on the position (in our case, because we use as starting point the transverse momentum distributions, these are positions in the plane transverse to the collision axis located at the collision point). Comparison of results from $p+p$ and central $Pb+Pb$ collisions supports our findings which seem to be supported by some experimental results.

Finally, note that results presented in \cite{WWln,RWWlnA} have been attributed to a discrete scale invariance. This  is a weaker form of the usual scale-invariance symmetry and results in some log-periodic corrections connected with a possible fractal structure of the process under consideration, cf., for example \cite{Sornette}. However, it turns out that such scale invariant functions also satisfy wave equations showing a self-similarity property. In both cases these functions exhibit log-periodic behavior. The functions discussed in our work, like $f\left(p_T\right)$, $T\left( p_T\right)$ and $T(r)$ are examples of this regularity.

\appendix

\section{Examples of mechanisms leading to Tsallis distribution}
\label{sec:A}

Tsallis distribution given by Eq. (\ref{TD}) is usually derived from Tsallis entropy, $S_q = - \frac{1}{(q-1)}\int dE f(E) \left[ f^{q-1}(E)-1\right]$, for probability density $f(E)$, constrained by condition that $ \int dE E f^q(E) =\langle E\rangle_q$ , and using the MaxEnt variational approach. It is worth to note that  Tsallis distribution can be also derived directly from the Shannon entropy, $S = - \int dE f(E)\ln[f(E)]$, supplied  with constraint:  $< \ln\left[1 - (1 - q)\frac{E}{T_0}\right] > = \frac{q-1}{2-q}$.

However, there are numerous examples of dynamical considerations, not based on MaxEnt approach, resulting in Eq. (\ref{TD}). Below we present their representative list (for details and references see \cite{WW_APPB}).
\begin{itemize}
\item {\bf Superstatistics.}~~  This is based on the observation that a gamma-like fluctuation of the scale parameter in exponential distribution results in the $q$-exponential Tsallis distribution, Eq. (\ref{TD}), with parameter $q$ defining the strength of fluctuations, $q = 1 + Var(T)/<T>^2$. In thermal approach it describes a non homogeneous heat bath in which temperature $T$ is different in different parts and fluctuates around some mean value, $T_0$.
\item {\bf Preferential attachment.}~~ In some systems (for example, in stochastic networks) one observes correlations of the preferential attachment type ("rich-get-richer") when the scale parameter $T_0$ depends on the variable under consideration. If $T_0\rightarrow T_0' (E) = T_0+(q-1)E$ then the probability distribution function, $f(E)$, is given by an equation $\frac{d f(E)}{d E} = - \frac{f(E)}{T'_0(E)}$, the solution of which is a Tsallis distribution, Eq. (\ref{TD}).
\item {\bf Multiplicative noise.}~~ In stochastic systems described by the Langevin equation,
\begin{equation}
\frac{d E}{dt} + \gamma(t) E = \xi(t), \label{eq:Le}
\end{equation}
with both the multiplicative, $\gamma(t)$, and additive, $\xi(t)$, noises, one gets for the distribution function $f(E)$ the Tsallis distribution, Eq. (\ref{TD}), with $n = 2 + \langle \gamma\rangle /Var(\gamma)$ and
$T(q) = \frac{(2-q)}{\langle \gamma\rangle}\left[ Cov(\xi,\gamma) + (q-1)\langle \xi\rangle\right]$.
\item {\bf Constrained systems.}~~For $N$ independent energies, $\left\{E_{i=1,\dots,N}\right\}$, each of them distributed according to the Boltzman distribution, their sum, $U = \sum_{i=1}^{N}E_i$, is distributed according to gamma distribution, $g_N(U) = 1/[ T_0 (N-1)](U/T_0)^{N-1} \exp(-U/T_0)$. If the available energy is limited, $U=N\alpha= const$, then the resulting {\it conditional probability} $f\left( E_i|U\right)= \frac{(N-1)}{N\alpha}\left( 1 - \frac{1}{N}\frac{E_i}{\alpha}\right)^{(N-2)}$ becomes Tsallis distribution, Eq. (\ref{TD}), with $p_T \to E_i$, $T_0 = \alpha N/(N-1)$ and $q =(N-3)/(N-2) < 1$. To get a Tsallis distribution with $q >1$ one would have to fluctuate $N$ according to Negative Binomial Distribution.
\item{\bf Statistical physics.}~~We end this list reminding how Tsallis distribution with $q<1$ arises from {\it statistical physics considerations}. Consider an isolated system with energy $U=const$ and with $\nu$
degrees of freedom. Choose single degree of freedom with energy $E$ (i.e., the remaining, or reservoir, energy is $E_r = U - E$). The number of states of the whole system is $\Omega(U-E)$ and probability that the energy of the chosen degree
of freedom is $E$ is $P(E) \propto \Omega(U-E)$. Expanding (slowly varying) $\ln \Omega(E)$ around $U$,
\begin{equation}
\ln \Omega(U-E) = \sum_{k=0}^{\infty} \frac{1}{k!}
\frac{\partial^{(k)} \ln \Omega}{\partial E_r^{(k)}}, \quad {\rm
with}\quad \beta = \frac{1}{T} \stackrel{def}{=}
\frac{\partial \ln \Omega\left(E_r\right)}{\partial
E_r},\label{eq:deriv}
\end{equation}
and (because $E<<U$) keeping only the two first terms one gets
\begin{equation}
\ln P(E) \propto \ln \Omega(E) \propto - \beta E,\quad {\rm or} \quad P(E) \propto \exp( - \beta E), \label{eq:BoltzmannD}
\end{equation}
i.e., a Boltzmann distribution (or $q=1$). On the other hand, because one usually expects that $\Omega\left(E_r\right) \propto \left(E_r/\nu\right)^{\alpha_1 \nu - \alpha_2}$ (where $\alpha_{1,2}$ are of the order of unity), one can write the full series for probability of choosing energy $E$:
\begin{eqnarray}
\!\!\! P(E)&\propto& \frac{\Omega(U-E)}{\Omega(U)} = \exp\left[ \sum_{k=0}^{\infty}\frac{(-1)^k}{k+1}\frac{1}{(\nu - 2)^k}(- \beta E)^{k+1}\right]=\nonumber\\
&=& \beta \frac{(\nu - 1)}{(\nu - 2)}\left(1 - \frac{1}{\nu - 2}\beta E\right)^{(\nu - 2)}.
\label{eq:statres}
\end{eqnarray}
This result, with $q=1 - 1/(\nu -2) \leq 1$, coincides with results from conditional probability.
\end{itemize}

There are still other possibilities details of which can be found in \cite{WW_APPB,WW_ChSF}.

\section{Comparison of log-oscillating transverse momenta $p_T$ and log-oscillating temperature $T$ }
\label{sec:B}

Note that the log-oscillating behavior of data on $p_T$ can be described in two ways, either by complex values of the nonextensivity parameter $q$ resulting in oscillatory dressing of the original Tsallis distribution (cf., for example, Eq. (51) in \cite{WW_ChSF}), or by log-oscilatory behavior of the scale parameter in this distribution, i.e., of the temperature $T$. Because in both cases they describe the same experimental data one can formally equate the corresponding distributions:
\begin{equation}
\left( 1 + \frac{p_T}{nT_0}\right)^{-m} \left[ 1 + a_1 f_1(p_T)\right] = \left\{ 1 + \frac{p_T}{nT_0\left[1 + a_2 f_2(p_T)\right]}\right\}^{-n}, \label{equal}
\end{equation}
where $f_1(p_T)$ and $f_2(p_T)$ are two log-periodic functions. Because in our case $m\simeq n$ and $\frac{p_T}{nT_0} >>1$ one can write that, approximately,
\begin{equation}
\left( \frac{p_T}{m T_0}\right)^{-m}\left[ 1 + a_1 f_1(p_T)\right] \simeq
\left\{ \frac{p_T}{mT_0\left[1 + a_2 f_2(p_T)\right]}\right\}^{-m}, \label{approx-equal}
\end{equation}
which means that in this approximation
\begin{equation}
\left[ 1 + a_1 f_1(p_T)\right] \simeq \left[ 1 + a_2 f_2(p_T)\right]^m. \label{eq}
\end{equation}
Because in our case $a_2 << 1$ and $\left| f_2\right| \leq 1$, one can write approximately that
\begin{equation}
\left[ 1 + a_2 f_2(p_T)\right]^m \simeq 1 + m a_2 f_2(p_T), \label{approx-eq}
\end{equation}
and, finally, one has that both log-periodic functions are approximately related in the following way:
\begin{equation}
a_1f_1(p_T) \simeq m a_2 f_2(p_T). \label{equivalence}
\end{equation}
Indeed, looking at the oscillations of $f\left( p_T\right)$ (cf., for example, Fig. 2 in \cite{WW_ChSF}) and $T\left( p_T\right)$ one realizes that the range of amplitudes of their oscillations is in reasonable agreement with this estimation.

\section*{Acknowledgements}

This research  was supported in part by the National Science Center (NCN) under contract 2016/22/M/ST2/00176.  We would like to thank warmly Dr Nicholas Keeley for reading the manuscript.

\end{document}